\documentclass[letter]{aa} 
\usepackage{graphicx}    
\usepackage[varg]{txfonts}  
\usepackage{upgreek} 
\usepackage{multirow} 
\usepackage[]{natbib}
\newcommand{\HII}{H\,{\sc ii}}

\newcommand{\CII}{C\,{\sc ii}}

\newcommand{\CI}{C\,{\sc i}}

\bibpunct{(}{)}{;}{a}{}{,}

\begin{document}

\title{Direct estimation of  electron density in the Orion Bar PDR\\
from mm-wave carbon recombination lines
\thanks{Based on observations obtained with the IRAM\,30\,m telescope supported by INSU/CNRS (France), MPG (Germany), and IGN (Spain).}}

  \author{S. Cuadrado\inst{\ref{inst1}}
  \and P. Salas\inst{\ref{inst2}}
  \and J. R. Goicoechea\inst{\ref{inst1}}\thanks{Corresponding author, \email{javier.r.goicoechea@csic.es}}
  \and J. Cernicharo\inst{\ref{inst1}}
  \and A. G. G. M. Tielens\inst{\ref{inst2}}
  \and A. B\'aez-Rubio\inst{\ref{inst3}}}
  
   \institute{
 Instituto de F\'isica Fundamental (IFF-CSIC). Calle Serrano 121-123, E28006 Madrid, Spain  \label{inst1}    
   \and Leiden Observatory, Leiden University, P.O. Box 9513, NL-2300 RA Leiden, The Netherlands \label{inst2}
     \and Centro de Astrobiolog\'ia (CSIC-INTA), Ctra. de Torrej\'on a Ajalvir, km 4, E28850 Torrej\'on de Ardoz, Madrid, Spain \label{inst3}}
   
   \date{Received 27 March 2019 / Accepted 16 April 2019}

%

\abstract{A significant fraction of the  molecular gas in star-forming regions is irradiated by stellar UV photons.  In these environments, the electron density \mbox{($n_{\rm e}$)}  plays a critical role in the gas dynamics, chemistry, and collisional excitation of certain molecules.} 
{We determine $n_{\rm e}$ in the prototypical strongly irradiated photodissociation region (PDR), the Orion Bar, from the detection of new millimeter-wave carbon recombination lines (mmCRLs)   and existing far-IR [$^{13}$\CII] hyperfine line observations.} 
{We detect 12 mmCRLs (including $\alpha$, $\beta$, and $\gamma$ transitions) observed with the IRAM\,30\,m telescope, at $\sim$\,25$''$ angular resolution, toward the H\,/\,H$_2$ dissociation front (DF) of the Bar. We also present a mmCRL emission cut across the PDR.} 
{These lines trace the \mbox{C$^+$\,/\,C\,/\,CO}  gas transition layer. As the much lower frequency carbon radio recombination lines, \mbox{mmCRLs} arise from neutral PDR gas and not from ionized gas in the adjacent \HII~region. This is readily seen
from their narrow line profiles \mbox{($\Delta v=2.6\pm0.4$~km\,s$^{-1}$)}
and line peak velocities \mbox{($v_{\rm LSR}=+10.7\pm0.2$~km\,s$^{-1}$)}. 
 Optically thin [$^{13}$\CII] hyperfine lines and molecular lines 
 -- emitted close to the DF by trace species such as reactive ions CO$^+$ and HOC$^+$ -- show the same line profiles.
We use non-LTE excitation models of [$^{13}$\CII] and mmCRLs 
and derive \mbox{$n_{\rm e}$\,$=$\,60\,--\,100~cm$^{-3}$}
and \mbox{$T_{\rm e}$\,$=$\,500\,--\,600~K} toward the DF.} {The inferred electron densities are high, up to an order of magnitude higher than 
previously thought. They provide a lower limit to the gas thermal pressure at the PDR edge without using molecular tracers. We obtain $P_{\rm th} \geq (2-4)\cdot10^8$\,cm$^{-3}$\,K
assuming that the electron abundance is
equal to or lower than the gas-phase elemental abundance of carbon.
Such elevated thermal pressures leave little room for magnetic pressure support and  agree with a scenario in which  the PDR photoevaporates.}


\titlerunning{The electron density in the Orion Bar}
\authorrunning{S. Cuadrado, P. Salas, J. R. Goicoechea et al.}

\maketitle


\section{Introduction}

Much of the mass and most of the volume occupied by molecular gas in  star-forming regions lies at low visual extinction  \citep[\mbox{$A_{\rm V}$\,$<$\,6}, e.g.,][]{Pety17}. 
This means that, in the vicinity of OB-type massive stars,  a significant fraction of the  molecular gas is irradiated by relatively intense UV photon fluxes \mbox{\citep[e.g.,][]{Goicoechea19}}.  The interface layers between the hot ionized gas and the cold molecular cloud are photodissociation regions \citep[PDRs;][]{Hollenbach_1999}. PDRs host the critical  
H$^+$\,/\,H\,/\,H$_2$ and \mbox{C$^+$\,/\,C\,/\,CO}  transition layers of the interstellar medium (ISM). \mbox{Far-UV} (FUV) photons with \mbox{$E<13.6$\,eV}  permeate molecular clouds,  ionizing  atoms, molecules, and dust grains of lower ionization potential (IPs).
One signature of \mbox{FUV-irradiated} gas is an ionization fraction,
defined as the abundance of electrons with respect  to H nuclei 
\mbox{($x_{\rm e}$\,=\,$n_{\rm e} / n_{\rm H}$)}, higher than about 
$10^{-6}$. Cold molecular cores shielded from external FUV radiation show much lower ionization fractions, $x_{\rm e} \lesssim 10^{-8}$, as the abundance of electrons is driven by the gentle flux of cosmic-ray particles rather than
penetrating stellar FUV photons 
\citep[][]{Guelin82, Caselli98, Maret07,Goicoechea_2009}.
 
 \begin{figure*}[ht]
 \centering
 \includegraphics[scale=0.35, angle=0]{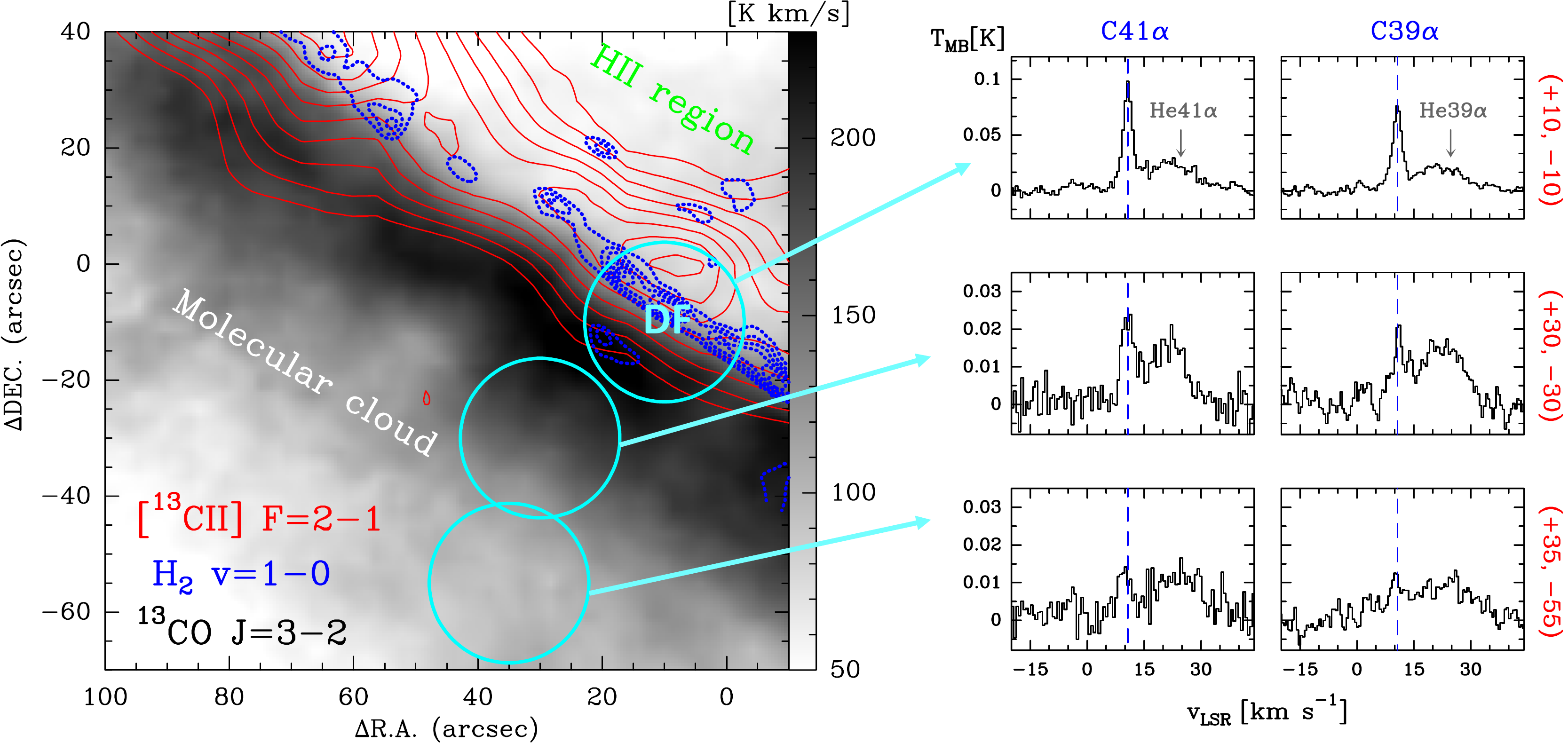}    
\caption{Detection of mmCRLs toward the Orion Bar PDR. 
\emph{Left:} Map of the \mbox{$^{13}$CO~$J$\,=\,3\,--\,2} integrated emission obtained with the IRAM\,30\,m telescope at a HPBW of 8$''$ (Cuadrado et al., in prep.). The blue dashed contours show the position of the H$_2$ dissociation front traced by the 
\mbox{H$_2$~$v$\,=\,1\,--\,0~$S$(1)} emission 
\citep[from 1.5 to 4.5\,$\cdot$\,10$^{-4}$~erg\,s$^{-1}$\,cm$^{-2}$\,sr$^{-1}$ in steps of 
0.5\,$\cdot$\,10$^{-4}$~erg\,s$^{-1}$\,cm$^{-2}$\,sr$^{-1}$; from][]{Walmsley_2000}.
Red contours show the  [$^{13}$\CII] ($^{2}P_{3/2}-^{2}P_{1/2}$, $F$\,=\,2--1) line emission at  1900.466~GHz  mapped with Herschel/HIFI at a 
HPBW of 12$''$
\citep[from 10 to 30~K\,km\,s$^{-1}$ in steps of 2.5~K\,km\,s$^{-1}$; from ][]{Goicoechea_2015}.
 \emph{Right:} C41$\alpha$ and C39$\alpha$ recombination lines detected with the IRAM\,30\,m telescope toward  three positions of the PDR. The cyan circles roughly represent
 the HPBW of our 3\,mm-wave observations.} 
 \label{fig:3pos}
\end{figure*}

Electrons
play a fundamental role in the chemistry and dynamics of the neutral interstellar gas
(meaning neutral atomic or molecular hydrogen, but not ionized). The  electron \mbox{density ($n_{\rm e}$)}  controls the preponderance of ion-neutral reactions, i.e., the main formation route for many ISM molecules \citep{Herbst73,Oppenheimer74}. The ionization fraction also controls the coupling of matter and magnetic fields. 
In addition, in high $x_{\rm e}$ environments, the large \mbox{cross sections} for inelastic collisions of electrons with certain high-dipole molecules such as HCN provide an additional source of rotational excitation \citep{Goldsmith17}. In these cases, the observed molecular line emission is no longer controlled by the most abundant collisional partner, H$_2$. Hence, the actual  value of $n_{\rm e}$ affects how gas densities are estimated.

A direct determination of $n_{\rm e}$ in molecular clouds  is usually not possible and we have to rely on indirect methods such as the observation of molecular ions and chemical modeling.
In FUV-illuminated environments, electrons are supplied by the photoionization of abundant elements such as carbon and sulfur (both with \mbox{IP\,$<$\,13.6~eV}), and also by the photoelectric effect on dust grains and polycyclic aromatic hydrocarbon (PAH) molecules \citep{Bakes94}.  In diffuse and translucent clouds, and at the \mbox{FUV-irradiated} edges of dense molecular clouds, most electrons arise from the ionization of carbon atoms.
Carbon recombination lines (CRLs),  in which a free \mbox{electron} \mbox{recombines} with 
carbon ions (C$^+$)
and cascades down from Rydberg electronic states to the ground while emitting photons, are expected to arise from neutral gas close to the \mbox{C$^+$\,/\,C\,/\,CO}  transition layer
\citep[e.g.,][]{Natta_1994} and  not from the hot \mbox{(electron temperature $T_{\rm e} \approx 10,000$~K)} ionized gas in the adjacent \HII~region. 
This is readily seen from the narrower CRLs profiles compared to the  broad  H and He recombination lines \citep[\mbox{$\Delta v \gtrsim$\,20\,km\,s$^{-1}$}, e.g.,][]{Churchwell78}.
This conclusion is also in line with photoionization models  where,
in \HII~regions, carbon is mainly in the form of higher ionization states \mbox{(e.g., C$^{++}$)} \citep[][]{Rubin91,Kaufman06}.

The \mbox{$^{2}P_{3/2}-^{2}P_{1/2}$} fine-structure emission of singly ionized carbon (IP\,$=$\,11.3\,eV), the famous [\CII]\,158\,$\upmu$m line, is bright
 and often shows an intensity linearly proportional to the C$^+$ column density
 \citep[the so-called effectively thin emission regime;][]{Goldsmith_2012}. However, the line reaches moderate opacities toward  bright and dense PDRs such as the Orion Bar \mbox{\citep[e.g.,][]{Ossenkopf13,Goicoechea_2015}}.
Carbon recombination lines are optically thin (see Sect.~4) with 
an intensity proportional to \mbox{$n_{\mathrm{e}}^2\,T_{\mathrm{e}}^{-2.5}$}. 
Although much fainter, \mbox{mmCRLs}  can be observed from ground-based telescopes
and can be used to infer $n_{\rm e}$ and $T_{\rm e}$
in \mbox{FUV-irradiated} neutral gas
\citep[][]{Pankonin78,Salgado17,Salas18}.
CRLs have  historically been detected at very low radio frequencies (e.g., at $\sim$\,43\,MHz for C539$\alpha$ or
$\sim$\,8.6\,GHz for C91$\alpha$). Pushing to higher frequencies 
(i.e., lower principal quantum numbers $n$) greatly improves  the angular resolution of the observation even with single-dish telescopes. This allows us to access much smaller spatial scales  and, potentially, to spatially resolve the narrow \mbox{C$^+$\,/\,C\,/\,CO}  gas transition layer. 

In this work we present
the detection of several \mbox{$\alpha$ ($\Delta n = 1$)}, \mbox{$\beta$ ($\Delta n = 2$)}, and \mbox{$\gamma$ ($\Delta n =3$)} mmCRLs  (C$n\Delta n$) observed 
from $\sim$\,85\,GHz to $\sim$\,115\,GHz toward
the strongly FUV-irradiated \mbox{($G_0$\,$\gtrsim$\,10$^4$)} PDR, the Orion Bar. This is a nearly edge-on interface of the Orion molecular cloud \mbox{(OMC-1)} with the ‘`Huygens'’ dense \HII~region, photoionized  by young massive stars in the
 Trapezium cluster \citep[e.g.,][]{Tielens_1993,Odell01,Goicoechea_2016,Pabst_2019}. 
Using the Effelsberg 100\,m telescope, \citet{Natta_1994} previously detected the C91$\alpha$ line toward several positions of the irradiated surface of  \mbox{OMC-1}. 
The same line was  mapped with the VLA along the Bar by \citet{Wyrowski_1997}.
They showed that the C91$\alpha$ emission basically coincides with
the emission in the \mbox{$v$\,$=$\,1--0 $S$(1)} line from vibrationally excited molecular hydrogen (H$_{2}^{*}$). Most  models
of the Bar use
\mbox{$n_e = 10$\,cm$^{-3}$} for the edge of the  PDR
\mbox{\citep[e.g.,][]{VanderTak_2012,van_der_Tak_2013}}.
This value implies relatively low gas densities
(\mbox{$n_{\rm H}$\,$\simeq$\,10$^5$~cm$^{-3}$})
and thermal pressures in the CRL emitting layers,
and through the PDR if the classical constant-density PDR model is adopted. 
The newly detected mmCRLs allow us to determine $n_{\rm e}$ and $T_{\rm e}$, and to independently estimate the gas thermal pressure.
This provides additional insights into the PDR structure and dynamics.


\section{Observations and data reduction}
We used the IRAM\,30\,m telescope at Pico Veleta (Sierra Nevada, Spain) to 
observe the Orion Bar in the mm band. We employed the E0 EMIR receiver (80\,GHz\,$-$\,116\,GHz)
and  fast Fourier transform spectrometer (FFTS) backend at 200\,kHz spectral resolution (0.7\,km\,s$^{-1}$ at 90 \,GHz). 
These observations are part of a complete  line survey  (80\,GHz\,$-$\,360\,GHz; \citealt{Cuadrado_2015,Cuadrado_2016,Cuadrado_2017})
toward a position close to the H$_2$ dissociation front (DF; the H\,/\,H$_2$ transition layer), almost coincident
with what is known as the  CO$^+$ emission peak \citep{Stoerzer_1995}.
Here we present results obtained from deep observations in the 3\,mm band  toward three positions across the PDR (see Fig.~\ref{fig:3pos}). Their offsets  with respect to  \mbox{$\mathrm{\alpha_{2000}=05^{h}\,35^{m}\,20.1^{s}\,}$}, \mbox{$\mathrm{\delta_{2000}=-\,05^{\circ}25'07.0''}$} are
 (+10$''$,\,$-$10$''$)\,$=$\,DF, (+30$''$,\,$-$30$''$), and
(+35$''$,\,$-$55$''$).
In order to avoid the extended emission from \mbox{OMC-1}, we employed the 
position switching observing procedure with a reference position at  offset ($-$600$''$,\,0$''$). 

The half power beam width (HPBW) at 3\,mm ranges from $\sim$\,31$''$ to $\sim$\,21$''$ (see Table~\ref{Table_CRRL_IRAM}).
We reduced and analyzed the data using the GILDAS software,\footnote{http://www.iram.fr/IRAMFR/GILDAS/} as described in \citet{Cuadrado_2015}. 
The rms noise obtained after 4\,h\,$-$\,5\,h integrations  is typically $\sim$\,1\,mK\,$-$\,5\,mK per resolution channel.
The antenna temperature, $T^{*}_{\rm A}$, was converted to the main beam temperature, $T_{\rm MB}$, through the \mbox{$T_{\rm MB}$ = $T^{*}_{\rm A}/ \upeta_{\rm MB}$} relation, where $\upeta_{\rm MB}$ is the antenna efficiency, which is defined as the ratio between main beam efficiency, $B_{\rm eff}$, and forward efficiency, $F_{\rm eff}$\footnote{http://www.iram.es/IRAMES/mainWiki/Iram30mEfficiencies}. All line intensities in figures and tables are in units of main beam temperature.

The intensities  of the C$n$$\alpha$ lines were extracted
from a two-Gaussian fit to each observed feature: one narrow Gaussian for the C$n$$\alpha$ lines, and a broader one for the He$n$$\alpha$ lines (see fits in Fig.~\ref{fig:CRRL_IRAM}). 
 With these fits we determined the contribution, $\approx40\,\%$,  of the He$n$$\alpha$ line wings to the observed emission at C$n$$\alpha$ velocities. We used this value to estimate the contribution of the putative He$n$$\beta$ and He$n$$\gamma$ line wings to the faint C$n$$\beta$ and C$n$$\gamma$ lines.
We conclude that the uncertainty (calibration and line overlap)
of our \mbox{mmCRL} intensities is $\approx 15\%$.
The resulting  \mbox{mmCRL} spectroscopic parameters are given in Table~\ref{Table_CRRL_IRAM}. 

We also made use of the [$^{13}$\CII] map taken by Herschel/HIFI toward \mbox{OMC-1} \citep{Goicoechea_2015}. 
We analyzed the strongest, yet \mbox{optically} thin, [$^{13}$\CII]
\mbox{$F$\,$=$2--1}  hyperfine emission component at 1900.466\,GHz 
(red contours in  Fig.~\ref{fig:3pos}). To make a comparison with the mmCRLs, we smoothed the map to an angular resolution of $\sim$25$''$ and extracted the [$^{13}$\CII] \mbox{($F$\,$=$2--1)} 
integrated line intensity, 20\,$\pm$\,3~K\,km\,s$^{-1}$, toward the DF.  

\section{Results} 

Figure~\ref{fig:3pos} shows the observed positions over a map of the optically thin \mbox{$^{13}$CO~($J$\,=\,3-2)} and
[$^{13}$\CII] \mbox{($^{2}P_{3/2}-^{2}P_{1/2}$, $F$\,=\,2--1)} emission lines
along the Bar.
We  detect 12 mmCRLs toward the DF: C42$\alpha$ to C38$\alpha$, C52$\beta$ to C48$\beta$, and  C60$\gamma$ to C59$\gamma$.
All lines are shown in Fig.~\ref{fig:CRRL_IRAM} of the Appendix. 
The emission from these lines gets fainter as we go  from the DF  to 
the more shielded molecular gas, thus mmCRLs trace the FUV-irradiated edge of the molecular cloud.
The C$n$$\alpha$ lines show an emission shoulder shifted by \mbox{$\simeq$\,$+$10~km\,s$^{-1}$}.
This feature is produced by He recombination lines (IP\,$=$\,24.6\,eV).
Helium lines do not arise from the neutral PDR; they are emitted from the 
surrounding ‘`Huygens'’ \HII~region and from  foreground layers of ionized gas
that extend all the way to the edge of Orion's Veil 
\citep[see, e.g.,][]{Rubin11,ODell17,Pabst_2019}.

\begin{figure}[t]
\centering
 \includegraphics[scale=0.77, angle=0]{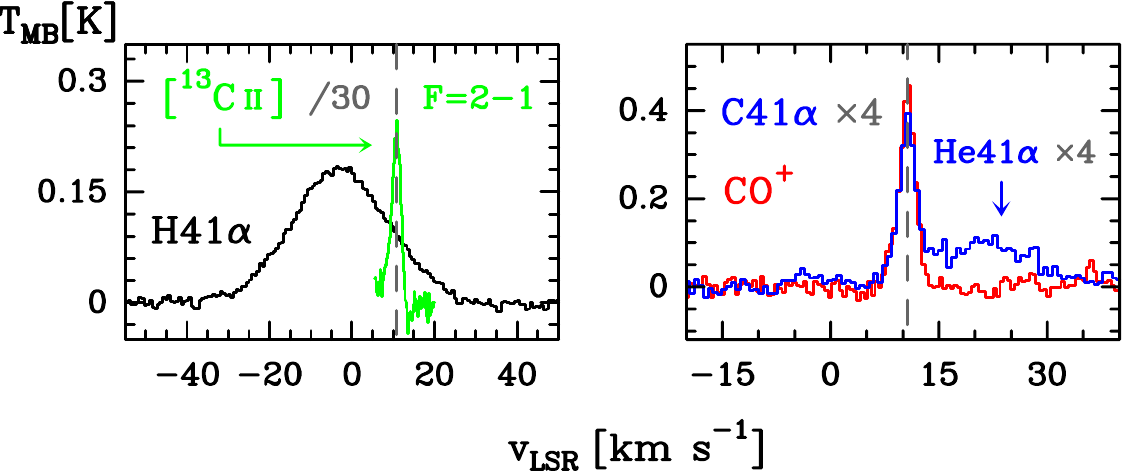}   
\caption{Line profiles toward the Orion Bar (DF position).
 \emph{Left:}~H41$\alpha$  with 
 \mbox{$\Delta v$\,$\simeq$\,26\,km\,s$^{-1}$} and \mbox{$v_{\rm LSR}$\,$\simeq$\,$-$4\,km\,s$^{-1}$}, and [$^{13}$\CII] (intensity divided by 30)
 with  \mbox{$\Delta v$\,$\simeq$\,2.6\,km\,s$^{-1}$}
 (the [$^{12}$\CII] emission has been blanked out).  \emph{Right:} CO$^+$ $N,F$\,=\,2,\,5/2\,--\,1,\,3/2  and C41$\alpha$  lines with $\Delta v$\,$\simeq$\,2.6\,km\,s$^{-1}$.
Dashed lines indicate the LSR velocity, \mbox{10.7 km s$^{-1}$}, of the PDR. 
}
 \label{fig:HC41a_CO+}
\end{figure}

 
The observed mmCRLs have line profiles that are very different from those of H and He recombination lines (\mbox{Fig.~\ref{fig:HC41a_CO+}}). The H and He recombination lines
show much broader line widths (\mbox{$\Delta v=10-30$~km\,s$^{-1}$}) produced by the high electron temperatures and pressures of the fully ionized gas. 
They peak at \mbox{$v_{\rm LSR}\,=\,-2$ to $-$11\,km\,s$^{-1}$},
consistent with ionized gas that flows toward the observer.
Carbon recombination lines, however, peak at \mbox{$v_{\rm LSR}=+10.7\pm0.2$~km\,s$^{-1}$}
and show narrow line profiles, \mbox{$\Delta v=2.6\pm0.4$~km\,s$^{-1}$}.
These values are nearly identical to those displayed by [$^{13}$\CII] and  by molecular lines observed toward the DF position
at comparable angular resolution \citep[e.g.,][]{Cuadrado_2015}. 
In particular,  mmCRLs and [$^{13}$\CII] line profiles are analogous to those of  HOC$^+$ and CO$^+$ (Fig.~\ref{fig:HC41a_CO+}). These reactive molecular ions  form by chemical reactions involving C$^+$ with H$_2$O and OH, respectively  \citep[e.g.,][]{Fuente03,Goico17}. 
Hence, they likely trace the same gas component.

For optically thin emission, line widths are determined by thermal broadening
($\propto$\,$\sqrt T_{\rm k}$) and by nonthermal broadening produced  
by gas turbulence and macroscopic motions in the  PDR. 
Adopting a nonthermal velocity dispersion\footnote{Calculated from detailed nonlocal radiative
transfer models of the molecular line emission toward the DF
\citep[][]{Goicoechea_2016,Goico17}.}  of  
$\sigma_{\rm nth}$\,=\,1.0\,$\pm$\,0.1\,km\,s$^{-1}$ 
($\Delta v_{\rm nth}$\,=\,2.355\,$\cdot$\,$\sigma_{\rm nth}$), the
observed mmCRL widths  imply a
beam-averaged gas temperature of \mbox{$T_{\rm k}\,=\,450^{+280}_{-300}$~K}.
The [\CII]\,158\,$\upmu$m line shows a broader line width, 
$\Delta v$\,$=$\,4.1\,$\pm$\,0.1~km\,s$^{-1}$, toward the DF. Because the line emission is moderately optically thick 
\citep[\mbox{$\tau_{\rm [CII]}$\,$\approx$\,1--2}; see][]{Ossenkopf13,Goicoechea_2015}, these line width differences are, at least in part, produced by opacity-broadening of the [\CII]\,158\,$\upmu$m line. 
However, \citet{Ossenkopf13} pointed out that opacity-broadening
alone does not fully explain the broader [\CII] line profile compared to [$^{13}$\CII].
These line width differences may suggest that, in comparison to [$^{13}$\CII] and mmCRLs,  the [\CII]\,158\,$\upmu$m emission has a significant contribution from hotter gas in the mostly atomic PDR \mbox{($x_{\rm H}$\,$>$\,$x_{\rm H_2}$)}, thus closer to
the  ionization front (the PDR\,/\,\HII~interface).

\section{Analysis}

Our \mbox{3\,mm-wave} observations have allowed us to  detect several $\alpha$, $\beta$, and $\gamma$ CRLs toward the Bar. 
The observed $n$ dependence of their line strengths is determined by the level populations. These can be modeled and used to derive $n_{\rm e}$ and $T_{\rm e}$ 
\citep[see theory in e.g.,][]{Walmsley_1982,Salgado17}. 



\begin{figure}
\centering
 \includegraphics[scale=0.95, angle=0]{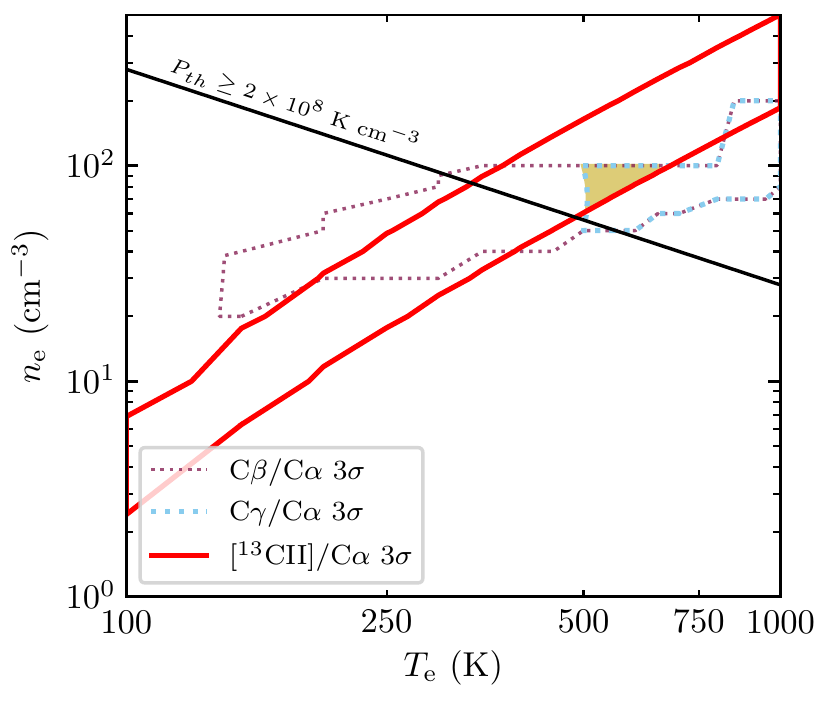}    
\caption{Constraints on $n_{\rm e}$ and $T_{\rm e}$ toward the Orion Bar DF 
from \mbox{non-LTE} excitation models that assume a path length of \mbox{$0.02 \leq l \leq 0.2$\,pc}. The colored area  shows the overlap region of models for different
line intensity ratios  (within a 3$\sigma$ uncertainty range in the observed ratios).}
\label{fig:gas_props_3sigma}
\end{figure}

Figure~\ref{fig:gas_props_3sigma} shows results of a grid of 
non-LTE\footnote{The observed mmCRL intensity ratios approach LTE for 
\mbox{$T_{\rm e}$\,$\gtrsim$\,500 K}.
Assuming LTE excitation  results in mmCRL intensities
brighter by $\lesssim25\%$. Hence, the estimated $n_{\rm e}$ in LTE  are $\lesssim25\%$ lower.}
 excitation models for $n_{\rm e}$ ranging from 1~cm$^{-3}$ to  500~cm$^{-3}$, and $T_{\rm e}$ ranging from 100\,K to 1000\,K.  Our models use non-LTE level populations computed by \citet{Salgado17}
without a background radiation field. 
Models assume that the observed lines are optically thin (for the conditions prevailing in the  Bar, we determine that the  opacity of the C$41\alpha$ line is
 \mbox{$\tau$\,$\simeq$\,$10^{-2}$)}.
Our models also compute the 
[$^{13}$\CII] \mbox{$^{2}P_{3-2}-^{2}P_{1/2}$} excitation,
and use the [$^{12}$C/$^{13}$C]\,$\simeq$\,67 isotopic abundance  ratio inferred  in Orion 
\citep{Langer_1984}.
The  colored area in Fig.~\ref{fig:gas_props_3sigma}  shows the best models fitting line intensity ratios that include 
all\footnote{The properties of the observed $\alpha$, $\beta$, and $\gamma$ carbon \mbox{recombination} lines vary slowly with $n$. In order to increase the \mbox{statistical} significance of  our comparison between models and observations, 
we used the inverse-variance weighted intensity averages of the observed
\mbox{C$n\alpha$ ($n$ from 38 to 42)}, \mbox{C$n\beta$ ($n$ from 48 to 52)}, and
\mbox{C$n\gamma$ ($n$ from 59 to 60)} lines.}
observed \mbox{$\alpha$, $\beta$, and $\gamma$} mmCRLs and [$^{13}$\CII]. 
The  black  line  shows where  the  gas  thermal  pressure
(\mbox{$P_{\rm th} = n_{\rm H}\cdot T_{\rm k}$}) is  2\,$\cdot$\,10$^{8}$\,cm$^{-3}$\,K.
To plot this line we assume \mbox{$x_{\rm e}$\,$=$\,$x_{\rm C^+}$\,$=$\,[C\,/\,H]}; in other words, all free electrons come from the ionization of carbon atoms, with an gas-phase abundance  of [C\,/\,H]\,=\,1.4\,$\cdot$\,10$^{-4}$ with respect to H nuclei in Orion \citep{Sofia_2004}.
Absolute line intensity predictions depend
on the assumed path-length $l$ along the line of sight. The
$\sim$\,25$''$ beam-averaged C$^+$ column density,  $N$(C$^+$),  
estimated from [$^{13}$\CII] is $N$(C$^+$)\,$\simeq$\,10$^{19}$~cm$^{-2}$ 
\citep{Goicoechea_2015}. Assuming a representative density of 
\mbox{$n_{\rm H}$\,$\simeq$\,10$^{5}$~cm$^{-3}$} in the atomic PDR 
\mbox{\citep{Tielens_1993}}, the inferred $N$(C$^+$) is equivalent to  $l$\,$\simeq$\,0.2~pc. This is consistent
with other estimations based on the infrared dust emission \citep[\mbox{$l$\,$=$\,0.28\,$\pm$\,0.06~pc},][]{Salgado16}.
If the gas density is a factor of ten higher \mbox{\citep[e.g.,][]{Labsch17}}
then  $l$\,$\simeq$\,0.02~pc. 

Our absolute intensity and line ratio models restrict $n_{\rm e}$ and $T_{\rm e}$ 
toward the DF position  to \mbox{60\,--\,100~cm$^{-3}$} and 
\mbox{500\,--\,600~K,} \mbox{respectively}. The
inferred electron temperatures in the colored area of Fig.~\ref{fig:gas_props_3sigma}
fall within the thermal line widths derived 
from the observed mmCRL profiles (see previous section).
 Assuming\footnote{Our inferred $n_{\rm H}$ and $P_{\rm th}$ values 
 are lower limits if mmCRLs arise from PDR gas layers where a significant fraction of carbon is not locked in C$^+$, and thus $x_{e} < 1.4\times10^{-4}$ 
 and $x_{e} > x$(C$^+$).}
\mbox{$x_{\rm e}$\,$\leq$\,1.4$\cdot$10$^{-4}$}, the derived electron densities
 are equivalent to gas densities of 
 \mbox{$n_{\rm H}$\,$\geq$\,(4\,--\,7)\,$\cdot$\,10$^5$~cm$^{-3}$}. Thus, gas
 thermal pressures of \mbox{$P_{\rm th}$\,$\geq$\,(2\,--\,4)$\cdot$10$^8$\,cm$^{-3}$\,K}
 toward the DF.

\section{Discussion and prospects}

Using mmCRL observations and models, we  inferred  
\mbox{$n_{\rm e}$\,$=$\,60--100~cm$^{-3}$} at the H\,/\,H$_2$ dissociation front of the Orion Bar PDR. These electron densities are higher than the 
 \mbox{$\simeq 10$~cm$^{-3}$} values typically used in 
 molecular excitation models of the region \mbox{\citep[e.g.,][]{VanderTak_2012,van_der_Tak_2013}}. In addition, by assuming \mbox{$x_{\rm e}$\,$\leq$\,1.4$\cdot$10$^{-4}$}, we  estimated
 a lower limit$^6$ to  $P_{\rm th}$ in the DF. The high inferred gas thermal pressures confirm earlier estimations based on the analysis of  ALMA images of the molecular gas emission  \citep{Goicoechea_2016,Goico17} and of Herschel observations of specific
   tracers of  the DF 
 \citep[e.g.,~\mbox{high-$J$~CO} and CH$^+$ rotational lines;][]{Nagy_2013,Joblin18}.
Nonstationary photoevaporating PDR models 
\mbox{\citep[e.g.,][]{Bertoldi_1996,Bron_2019}} 
predict such high pressures in PDRs.
In these time-dependent models, the strong stellar FUV field heats, compresses, and gradually evaporates the molecular cloud edge if the pressure of the surrounding medium (the adjacent \HII~region) is not significantly higher. 
The derived thermal pressure toward the DF, 
\mbox{$P_{\rm th}$\,$\gtrsim$\,2$\cdot$10$^8$\,cm$^{-3}$\,K},
is indeed higher than that of the ionized gas at the ionization front
\citep[\mbox{$\approx$\,6$\cdot$10$^7$\,cm$^{-3}$\,K},][]{Walmsley_2000} and,
in contrast to previous indirect studies of the pressure in the Bar
\citep{Pellegrini09}, leaves little room for magnetic pressure support. 
This conclusion is in line with the relatively modest plane-of-the-sky magnetic field strength
 reported from far-IR polarimetric observations   with
SOFIA/HAWC+ \citep{Chuss19}.




Unfortunately, the $\sim$25$''$ resolution of our single-dish  observations 
does not allow us to spatially resolve 
the [$^{13}$\CII] and mmCRLs emitting layers. 
We note that $A_{\rm V}$\,=\,1, roughly the width of the H\,/\,H$_2$ transition layer, implies 3.2$''$\,$-$\,1.6$''$  for $n_{\rm H}$\,=\,10$^5$ and
10$^6$\,cm$^{-3}$, respectively.
The $\sim$\,10$''$ resolution VLA map of the C91$\alpha$ line 
\citep{Wyrowski_1997}
shows that the C$^+$ gas layer seen in this CRL is spatially coincident with the IR emission from H$_{2}^{*}$ that traces the H\,/\,H$_2$ dissociation front
(shown in Fig.~\ref{fig:3pos}).
 This result is somewhat surprising because constant-density stationary  PDR models have long predicted
that the \mbox{C$^+$\,/\,C\,/\,CO}  transition in the Bar should be located deeper inside the  cloud, and separated from the DF
by several arcsec  \mbox{\citep[e.g.,][]{Tielens_1993}}. 
In addition, single-dish observations show that the \mbox{[\CI]\,492\,GHz}  emission spatially correlates with that of 
\mbox{$^{13}$CO~($J$\,=\,2--1)} \citep{Tauber95}. This suggests that 
the classical  \mbox{C$^+$\,/\,C\,/\,CO} sandwich
structure of a PDR may not be discernible, or even exist, in the sense that there would be no layer in the Bar where neutral atomic carbon is the most abundant carbon reservoir.
Indeed, ALMA images of the Bar at $\approx$\,1$''$ resolution show that there is also no appreciable offset between the H$_{2}^{*}$ emission and the edge of the HCO$^+$ and CO emission  \citep{Goicoechea_2016}. 
All these new observations thus suggest that we still do not fully understand the properties and exact location  of the \mbox{C$^+$\,/\,C\,/\,CO}  transition in \mbox{interstellar} clouds. 
 
In this work we  provided evidence that the electron density at the edge of the Orion Bar PDR is
quite high, and this may have consequences for the coupling of matter with the magnetic field and the excitation of certain molecules. Much higher resolution ALMA observations
of mmCRLs and of  neutral atomic carbon \mbox{[\CI] fine-structure} lines are clearly needed to spatially resolve these critical interface layers
of the ISM.

\begin{acknowledgements}
We thank the Spanish  MICIU for funding  support under grant AYA2017-85111-P
and the   ERC for support under grant ERC-2013-Syg-610256-NANOCOSMOS.
A.B.-R. also acknowledges support by the MICIU  and FEDER funding under grants ESP2015-65597-C4-1-R and ESP2017-86582-C4-1-R.
P.S. and  A.G.G.M.T. acknowledge financial support from the Dutch Science Organisation  through TOP grant 614.001.351.

\end{acknowledgements}

\bibliographystyle{aa}
\bibliography{references}

\begin{thebibliography}{49}
\expandafter\ifx\csname natexlab\endcsname\relax\def\natexlab#1{#1}\fi

\bibitem[{{Andree-Labsch} {et~al.}(2017){Andree-Labsch}, {Ossenkopf-Okada}, \&
  {R{\"o}llig}}]{Labsch17}
{Andree-Labsch}, S., {Ossenkopf-Okada}, V., \& {R{\"o}llig}, M. 2017, \aap,
  598, A2

\bibitem[{{Bakes} \& {Tielens}(1994)}]{Bakes94}
{Bakes}, E.~L.~O. \& {Tielens}, A.~G.~G.~M. 1994, \apj, 427, 822

\bibitem[{{Bertoldi} \& {Draine}(1996)}]{Bertoldi_1996}
{Bertoldi}, F. \& {Draine}, B.~T. 1996, \apj, 458, 222

\bibitem[{{Bron} {et~al.}(2019){Bron}, {Ag{\'u}ndez}, {Goicoechea}, \&
  {Cernicharo}}]{Bron_2019}
{Bron}, E., {Ag{\'u}ndez}, M., {Goicoechea}, J.~R., \& {Cernicharo}, J. 2019,
  arXiv e-prints

\bibitem[{{Caselli} {et~al.}(1998){Caselli}, {Walmsley}, {Terzieva}, \&
  {Herbst}}]{Caselli98}
{Caselli}, P., {Walmsley}, C.~M., {Terzieva}, R., \& {Herbst}, E. 1998, \apj,
  499, 234

\bibitem[{{Churchwell} {et~al.}(1978){Churchwell}, {Smith}, {Mathis}, {Mezger},
  \& {Huchtmeier}}]{Churchwell78}
{Churchwell}, E., {Smith}, L.~F., {Mathis}, J., {Mezger}, P.~G., \&
  {Huchtmeier}, W. 1978, \aap, 70, 719

\bibitem[{{Chuss} {et~al.}(2019){Chuss}, {Andersson}, {Bally}, {Dotson},
  {Dowell}, {Guerra}, {Harper}, {Houde}, {Jones}, {Lazarian}, {Lopez
  Rodriguez}, {Michail}, {Morris}, {Novak}, {Siah}, {Staguhn}, {Vaillancourt},
  {Volpert}, {Werner}, {Wollack}, {Benford}, {Berthoud}, {Cox}, {Crutcher},
  {Dale}, {Fissel}, {Goldsmith}, {Hamilton}, {Hanany}, {Henning}, {Looney},
  {Moseley}, {Santos}, {Stephens}, {Tassis}, {Trinh}, {Van Camp},
  {Ward-Thompson}, \& {HAWC + Science Team}}]{Chuss19}
{Chuss}, D.~T., {Andersson}, B.-G., {Bally}, J., {et~al.} 2019, \apj, 872, 187

\bibitem[{{Cuadrado} {et~al.}(2017){Cuadrado}, {Goicoechea}, {Cernicharo},
  {Fuente}, {Pety}, \& {Tercero}}]{Cuadrado_2017}
{Cuadrado}, S., {Goicoechea}, J.~R., {Cernicharo}, J., {et~al.} 2017, \aap,
  603, A124

\bibitem[{{Cuadrado} {et~al.}(2015){Cuadrado}, {Goicoechea}, {Pilleri},
  {Cernicharo}, {Fuente}, \& {Joblin}}]{Cuadrado_2015}
{Cuadrado}, S., {Goicoechea}, J.~R., {Pilleri}, P., {et~al.} 2015, \aap, 575,
  A82

\bibitem[{{Cuadrado} {et~al.}(2016){Cuadrado}, {Goicoechea}, {Roncero},
  {Aguado}, {Tercero}, \& {Cernicharo}}]{Cuadrado_2016}
{Cuadrado}, S., {Goicoechea}, J.~R., {Roncero}, O., {et~al.} 2016, \aap, 596,
  L1

\bibitem[{{Fuente} {et~al.}(2003){Fuente}, {Rodr\'{\i}guez-Franco},
  {Garc\'{\i}a-Burillo}, {Mart\'{\i}n-Pintado}, \& {Black}}]{Fuente03}
{Fuente}, A., {Rodr\'{\i}guez-Franco}, A., {Garc\'{\i}a-Burillo}, S.,
  {Mart\'{\i}n-Pintado}, J., \& {Black}, J.~H. 2003, \aap, 406, 899

\bibitem[{{Goicoechea} {et~al.}(2017){Goicoechea}, {Cuadrado}, {Pety}, {Bron},
  {Black}, {Cernicharo}, {Chapillon}, {Fuente}, \& {Gerin}}]{Goico17}
{Goicoechea}, J.~R., {Cuadrado}, S., {Pety}, J., {et~al.} 2017, \aap, 601, L9

\bibitem[{{Goicoechea} {et~al.}(2016){Goicoechea}, {Pety}, {Cuadrado},
  {Cernicharo}, {Chapillon}, {Fuente}, {Gerin}, {Joblin}, {Marcelino}, \&
  {Pilleri}}]{Goicoechea_2016}
{Goicoechea}, J.~R., {Pety}, J., {Cuadrado}, S., {et~al.} 2016, \nat, 537, 207

\bibitem[{{Goicoechea} {et~al.}(2009){Goicoechea}, {Pety}, {Gerin},
  {Hily-Blant}, {Le Bourlot}, {Le Bourlot}, {Le Bourlot}, {Le Bourlot}, \& {Le
  Bourlot}}]{Goicoechea_2009}
{Goicoechea}, J.~R., {Pety}, J., {Gerin}, M., {et~al.} 2009, \aap, 498, 771

\bibitem[{{Goicoechea} {et~al.}(2019){Goicoechea}, {Santa-Maria}, {Bron},
  {Teyssier}, {Marcelino}, {Cernicharo}, \& {Cuadrado}}]{Goicoechea19}
{Goicoechea}, J.~R., {Santa-Maria}, M.~G., {Bron}, E., {et~al.} 2019, \aap,
  622, A91

\bibitem[{{Goicoechea} {et~al.}(2015){Goicoechea}, {Teyssier}, {Etxaluze},
  {Goldsmith}, {Ossenkopf}, {Gerin}, {Bergin}, {Black}, {Cernicharo},
  {Cuadrado}, {Encrenaz}, {Falgarone}, {Fuente}, {Hacar}, {Lis}, {Marcelino},
  {Melnick}, {M{\"u}ller}, {Persson}, {Pety}, {R{\"o}llig}, {Schilke}, {Simon},
  {Snell}, \& {Stutzki}}]{Goicoechea_2015}
{Goicoechea}, J.~R., {Teyssier}, D., {Etxaluze}, M., {et~al.} 2015, \apj, 812,
  75

\bibitem[{{Goldsmith} \& {Kauffmann}(2017)}]{Goldsmith17}
{Goldsmith}, P.~F. \& {Kauffmann}, J. 2017, \apj, 841, 25

\bibitem[{{Goldsmith} {et~al.}(2012){Goldsmith}, {Langer}, {Pineda}, \&
  {Velusamy}}]{Goldsmith_2012}
{Goldsmith}, P.~F., {Langer}, W.~D., {Pineda}, J.~L., \& {Velusamy}, T. 2012,
  \apjs, 203, 13

\bibitem[{{Guelin} {et~al.}(1982){Guelin}, {Langer}, \& {Wilson}}]{Guelin82}
{Guelin}, M., {Langer}, W.~D., \& {Wilson}, R.~W. 1982, \aap, 107, 107

\bibitem[{{Herbst} \& {Klemperer}(1973)}]{Herbst73}
{Herbst}, E. \& {Klemperer}, W. 1973, \apj, 185, 505

\bibitem[{{Hollenbach} \& {Tielens}(1999)}]{Hollenbach_1999}
{Hollenbach}, D.~J. \& {Tielens}, A.~G.~G.~M. 1999, RevModPhys., 71, 173

\bibitem[{{Joblin} {et~al.}(2018){Joblin}, {Bron}, {Pinto}, {Pilleri}, {Le
  Petit}, {Gerin}, {Le Bourlot}, {Fuente}, {Berne}, {Goicoechea}, {Habart},
  {K{\"o}hler}, {Teyssier}, {Nagy}, {Montillaud}, {Vastel}, {Cernicharo},
  {R{\"o}llig}, {Ossenkopf-Okada}, \& {Bergin}}]{Joblin18}
{Joblin}, C., {Bron}, E., {Pinto}, C., {et~al.} 2018, \aap, 615, A129

\bibitem[{{Kaufman} {et~al.}(2006){Kaufman}, {Wolfire}, \&
  {Hollenbach}}]{Kaufman06}
{Kaufman}, M.~J., {Wolfire}, M.~G., \& {Hollenbach}, D.~J. 2006, \apj, 644, 283

\bibitem[{{Langer} {et~al.}(1984){Langer}, {Graedel}, {Frerking}, \&
  {Armentrout}}]{Langer_1984}
{Langer}, W.~D., {Graedel}, T.~E., {Frerking}, M.~A., \& {Armentrout}, P.~B.
  1984, \apj, 277, 581

\bibitem[{{Maret} \& {Bergin}(2007)}]{Maret07}
{Maret}, S. \& {Bergin}, E.~A. 2007, \apj, 664, 956

\bibitem[{{Nagy} {et~al.}(2013){Nagy}, {Van der Tak}, {Ossenkopf}, {Gerin}, {Le
  Petit}, {Le Bourlot}, {Black}, {Goicoechea}, {Joblin}, {R{\"o}llig}, \&
  {Bergin}}]{Nagy_2013}
{Nagy}, Z., {Van der Tak}, F.~F.~S., {Ossenkopf}, V., {et~al.} 2013, \aap, 550,
  A96

\bibitem[{{Natta} {et~al.}(1994){Natta}, {Walmsley}, \& {Tielens}}]{Natta_1994}
{Natta}, A., {Walmsley}, C.~M., \& {Tielens}, A.~G.~G.~M. 1994, \apj, 428, 209

\bibitem[{{O'Dell}(2001)}]{Odell01}
{O'Dell}, C.~R. 2001, \araa, 39, 99

\bibitem[{{O'Dell} {et~al.}(2017){O'Dell}, {Kollatschny}, \&
  {Ferland}}]{ODell17}
{O'Dell}, C.~R., {Kollatschny}, W., \& {Ferland}, G.~J. 2017, \apj, 837, 151

\bibitem[{{Oppenheimer} \& {Dalgarno}(1974)}]{Oppenheimer74}
{Oppenheimer}, M. \& {Dalgarno}, A. 1974, \apj, 192, 29

\bibitem[{{Ossenkopf} {et~al.}(2013){Ossenkopf}, {R{\"o}llig}, {Neufeld},
  {Pilleri}, {Lis}, {Fuente}, {van der Tak}, \& {Bergin}}]{Ossenkopf13}
{Ossenkopf}, V., {R{\"o}llig}, M., {Neufeld}, D.~A., {et~al.} 2013, \aap, 550,
  A57

\bibitem[{{Pabst} {et~al.}(2019){Pabst}, {Higgins}, {Goicoechea}, {Teyssier},
  {Berne}, {Chambers}, {Wolfire}, {Suri}, {Guesten}, {Stutzki}, {Graf},
  {Risacher}, \& {Tielens}}]{Pabst_2019}
{Pabst}, C., {Higgins}, R., {Goicoechea}, J.~R., {et~al.} 2019, \nat, 565, 618

\bibitem[{{Pankonin} \& {Walmsley}(1978)}]{Pankonin78}
{Pankonin}, V. \& {Walmsley}, C.~M. 1978, \aap, 67, 129

\bibitem[{{Pellegrini} {et~al.}(2009){Pellegrini}, {Baldwin}, {Ferland},
  {Shaw}, {Heathcote}, {Heathcote}, \& {Heathcote}}]{Pellegrini09}
{Pellegrini}, E.~W., {Baldwin}, J.~A., {Ferland}, G.~J., {et~al.} 2009, \apj,
  693, 285

\bibitem[{{Pety} {et~al.}(2017){Pety}, {Guzm{\'a}n}, {Orkisz}, {Liszt},
  {Gerin}, {Bron}, {Bardeau}, {Goicoechea}, {Gratier}, {Le Petit}, {Levrier},
  {{\"O}berg}, {Roueff}, \& {Sievers}}]{Pety17}
{Pety}, J., {Guzm{\'a}n}, V.~V., {Orkisz}, J.~H., {et~al.} 2017, \aap, 599, A98

\bibitem[{{Rubin} {et~al.}(1991){Rubin}, {Simpson}, {Haas}, \&
  {Erickson}}]{Rubin91}
{Rubin}, R.~H., {Simpson}, J.~P., {Haas}, M.~R., \& {Erickson}, E.~F. 1991,
  \apj, 374, 564

\bibitem[{{Rubin} {et~al.}(2011){Rubin}, {Simpson}, {O'Dell}, {McNabb},
  {Colgan}, {Zhuge}, {Ferland}, \& {Hidalgo}}]{Rubin11}
{Rubin}, R.~H., {Simpson}, J.~P., {O'Dell}, C.~R., {et~al.} 2011, \mnras, 410,
  1320

\bibitem[{{Salas} {et~al.}(2018){Salas}, {Oonk}, {van Weeren}, {Wolfire},
  {Emig}, {Toribio}, {R{\"o}ttgering}, \& {Tielens}}]{Salas18}
{Salas}, P., {Oonk}, J.~B.~R., {van Weeren}, R.~J., {et~al.} 2018, \mnras, 475,
  2496

\bibitem[{{Salgado} {et~al.}(2016){Salgado}, {Bern{\'e}}, {Adams}, {Herter},
  {Keller}, \& {Tielens}}]{Salgado16}
{Salgado}, F., {Bern{\'e}}, O., {Adams}, J.~D., {et~al.} 2016, \apj, 830, 118

\bibitem[{{Salgado} {et~al.}(2017){Salgado}, {Morabito}, {Oonk}, {Salas},
  {Toribio}, {R{\"o}ttgering}, \& {Tielens}}]{Salgado17}
{Salgado}, F., {Morabito}, L.~K., {Oonk}, J.~B.~R., {et~al.} 2017, \apj, 837,
  141

\bibitem[{{Sofia} {et~al.}(2004){Sofia}, {Lauroesch}, {Meyer}, \&
  {Cartledge}}]{Sofia_2004}
{Sofia}, U.~J., {Lauroesch}, J.~T., {Meyer}, D.~M., \& {Cartledge}, S.~I.~B.
  2004, \apj, 605, 272

\bibitem[{{Stoerzer} {et~al.}(1995){Stoerzer}, {Stutzki}, \&
  {Sternberg}}]{Stoerzer_1995}
{Stoerzer}, H., {Stutzki}, J., \& {Sternberg}, A. 1995, \aap, 296, L9

\bibitem[{{Tauber} {et~al.}(1995){Tauber}, {Lis}, {Keene}, {Schilke}, \&
  {B\"{u}ttgenbach}}]{Tauber95}
{Tauber}, J.~A., {Lis}, D.~C., {Keene}, J., {Schilke}, P., \&
  {B\"{u}ttgenbach}, T.~H. 1995, \aap, 297, 567

\bibitem[{{Tielens} {et~al.}(1993){Tielens}, {Meixner}, {van der Werf},
  {Bregman}, {Tauber}, {Stutzki}, \& {Rank}}]{Tielens_1993}
{Tielens}, A.~G.~G.~M., {Meixner}, M.~M., {van der Werf}, P.~P., {et~al.} 1993,
  Science, 262, 86

\bibitem[{{van der Tak} {et~al.}(2013){van der Tak}, {Nagy}, {Ossenkopf},
  {Makai}, {Black}, {Faure}, {Gerin}, \& {Bergin}}]{van_der_Tak_2013}
{van der Tak}, F.~F.~S., {Nagy}, Z., {Ossenkopf}, V., {et~al.} 2013, \aap, 560,
  A95

\bibitem[{{van der Tak} {et~al.}(2012){van der Tak}, {Ossenkopf}, {Nagy},
  {Faure}, {R{\"o}llig}, \& {Bergin}}]{VanderTak_2012}
{van der Tak}, F.~F.~S., {Ossenkopf}, V., {Nagy}, Z., {et~al.} 2012, \aap, 537,
  L10

\bibitem[{{Walmsley} {et~al.}(2000){Walmsley}, {Natta}, {Oliva}, \&
  {Testi}}]{Walmsley_2000}
{Walmsley}, C.~M., {Natta}, A., {Oliva}, E., \& {Testi}, L. 2000, \aap, 364,
  301

\bibitem[{{Walmsley} \& {Watson}(1982)}]{Walmsley_1982}
{Walmsley}, C.~M. \& {Watson}, W.~D. 1982, \apj, 260, 317

\bibitem[{{Wyrowski} {et~al.}(1997){Wyrowski}, {Schilke}, {Hofner}, \&
  {Walmsley}}]{Wyrowski_1997}
{Wyrowski}, F., {Schilke}, P., {Hofner}, P., \& {Walmsley}, C.~M. 1997, \apjl,
  487, L171

\end{thebibliography}

\appendix

\section{Tables and figures}
 
\begin{figure*}
\vspace*{0.5cm}
\centering 
\includegraphics[scale=0.7, angle=0]{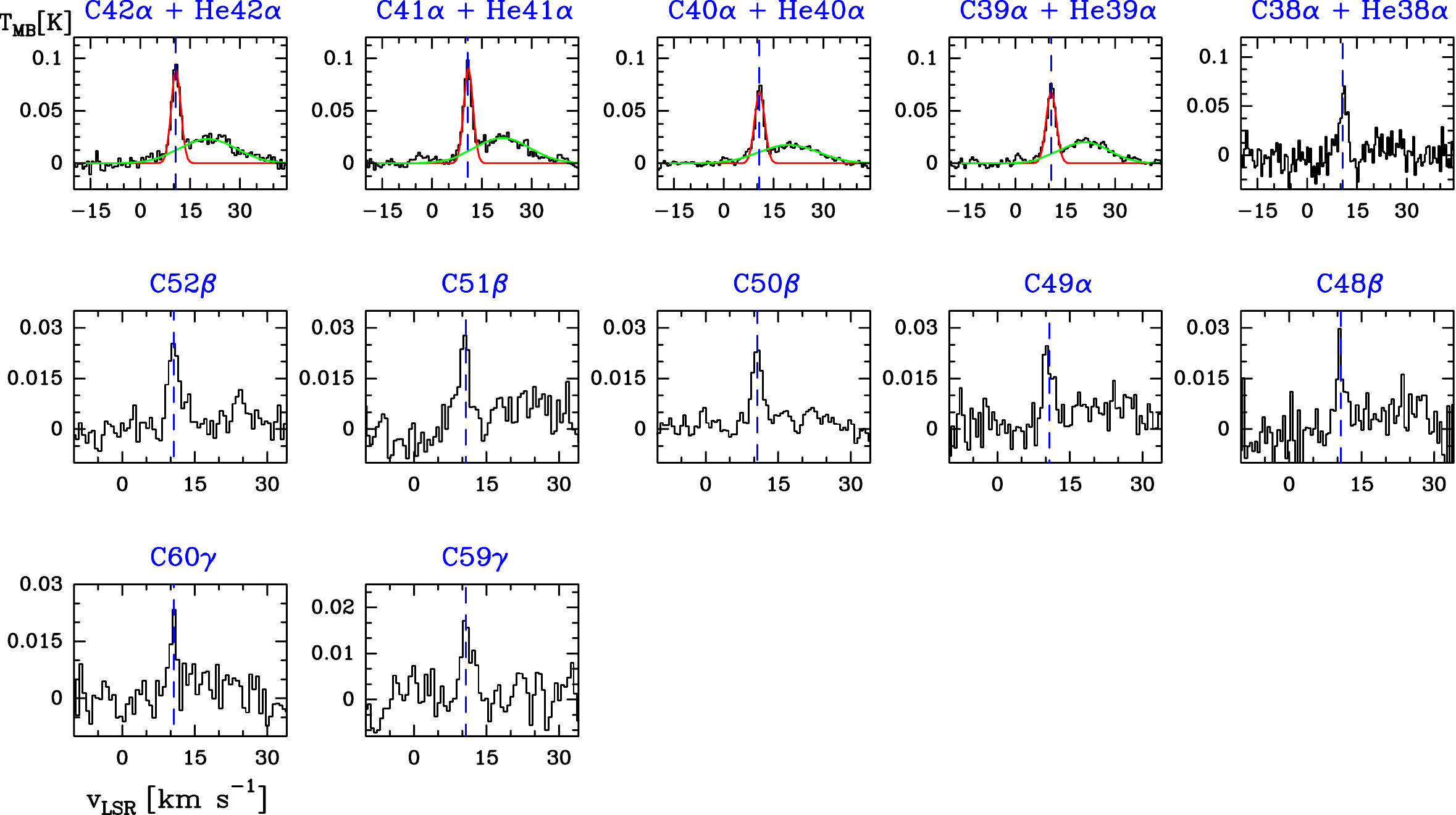}    
\caption{Carbon recombination  lines between 80\,GHz and 116\,GHz
detected with the IRAM\,30\,m telescope toward the Orion Bar (DF position). The dashed lines indicate the LSR
velocity \mbox{(10.7 km s$^{-1}$)} of the molecular gas in the PDR.
The red and green curves show Gaussian fits to the C$n \alpha$ and He$n \alpha$
lines, respectively. We use these fits to determine the contribution of He recombination
 line wings to the emission observed in the velocity range of the C$n \beta$ and C$n \gamma$ lines. 
We note the different abscissa and ordinate axis scales.}\label{fig:CRRL_IRAM}
\end{figure*}

\begin{table*}[!ht] 
\begin{center}
\vspace*{2cm}
\caption{Line spectroscopic parameters obtained from Gaussian fits to
the observed mmCRLs (see Sect.~2).}  \label{Table_CRRL_IRAM} 
\vspace{0.15cm}         
       \begin{tabular}{c c c c c c c c@{\vrule height 10pt depth 5pt width 0pt}}    
       \hline\hline       
       
Line &   Frequency  &   $\displaystyle{\int} T_{\rm MB}$d$v^{\,a,b}$  &   $v_{\rm LSR}$$^{\,b}$ &  $\Delta v$$^{\,b}$ &  $T_{\rm MB}^{\,a}$ & $S/N$$^{\,c}$ & HPBW$^{\,d}$ \rule[-0.3cm]{0cm}{0.8cm}\ \\     
      
   & [MHz] & [mK km s$^{-1}$]  & [km s$^{-1}$] & [km s$^{-1}$]  & [mK] & & [arcsec] \\  
          
 \hline
 
 C42$\alpha$   &      85731.14   &    226.8\,(10.5)  &    10.6\,(0.1)  & 2.6\,(0.1)   &  83.1   & 21  &  28.7 \\    
 C41$\alpha$   &      92080.35   &    248.9\,(14.2)  &    10.8\,(0.1)  & 2.7\,(0.1)   &  85.6   & 17  &  26.7 \\ 
 C40$\alpha$   &      99072.36   &    172.6\,(7.2)   &    10.7\,(0.1)  & 2.5\,(0.1)   &  63.6   & 23  &  24.8 \\ 
 C39$\alpha$   &     106790.61   &    190.9\,(13.3)  &    10.7\,(0.1)  & 2.9\,(0.2)   &  53.5   & 12  &  23.0 \\   
 C38$\alpha$   &     115331.91   &    163.9\,(19.6)  &    10.9\,(0.2)  & 2.4\,(0.3)   &  65.4   &  5  &  21.3 \\  

 \hline

 C52$\beta$    &      88449.80   &   53.8\,(9.4)       &  10.7\,(0.2)  & 2.9\,(0.5)   &  24.5  &  6  & 27.8  \\     
 C51$\beta$    &      93654.02   &   55.2\,(8.3)       &  10.5\,(0.2)  & 2.9\,(0.6)   &  24.8  &  6  & 26.3  \\   
 C50$\beta$    &      99274.72   &   47.7\,(6.0)       &  10.7\,(0.1)  & 2.7\,(0.3)   &  23.6  &  8  & 24.8  \\   
 C49$\beta$    &     105354.40   &   42.4\,(8.5)       &  10.6\,(0.2)  & 2.6\,(0.5)   &  21.5  &  4  & 23.3  \\  
 C48$\beta$    &     111940.89   &   36.9\,(11.0)      &  10.6\,(0.2)  & 2.3\,(0.6)   &  21.8  &  4  & 22.0  \\   

 \hline                                                         
 C60$\gamma$   &      84956.76   &   27.8\,(8.2)     &  10.5\,(0.2)  & 1.7\,(0.5)   &  22.3  &  5  & 29.0   \\     
 C59$\gamma$   &      89243.05   &   35.1\,(8.3)     &  10.9\,(0.3)  & 3.0\,(0.6)   &  15.4  &  4  & 27.6   \\

\hline                                                                                                                   
\end{tabular}   
\tablefoot{$^a$~Intensities in main beam temperature (in units of mK).
     $^b$~Parentheses indicate the uncertainty obtained by the Gaussian fitting routine.
      $^c$~Signal-to-noise ratio with respect to the peak line temperature in
      velocity resolution channels of 0.7~km\,s$^{-1}$.
      $^d$~The half power beam width (HPBW) of the IRAM\,30\,m telescope is well
      described by \mbox{HPBW[arcsec] $\approx$ 2460/Frequency[GHz]}.}
\end{center}              
\end{table*}  
  
\end{document}